\documentclass[preprint,12pt,authoryear]{elsarticle}

\usepackage{graphicx}
\usepackage{url}
\usepackage{apalike}
\usepackage{natbib}
\usepackage[colorinlistoftodos]{todonotes}
\usepackage[hidelinks]{hyperref}
\usepackage[figurename= Fig.]{caption}

\usepackage{scrextend} 

\journal{arxiv.org}

\begin{document}
\begin{frontmatter}

\title{Electronic schematic for bio-plausible dopamine neuromodulation of eSTDP and iSTDP}

\author[inst1]{Max Talanov}
\ead{max.talanov@gmail.com}
\author[inst1]{Yuriy Gerasimov}
\ead{yurger2009@gmail.com}
\author[inst1,inst2]{Victor Erokhin}
\ead{victor.erokhin@fis.unipr.it}

\address[inst1]{KFU, Russia.}
\address[inst2]{CNR-IMEM, Italy.}


\begin{abstract}

  In this technical report we present novel results of the dopamine bio-plausible neuromodulation excitatory (eSTDP) and inhibitory (iSTDP) learning. 
  We present the principal schematic for the neuromodulation of D1 and D2 receptors of dopamine, wiring schematic for both cases as well as the simulatory experiments results done in LTSpice. 
\end{abstract}
\end{frontmatter}

\section{The experimental set-up}

In our previous technical report we presented the electronic schematic to generate complex learning impulses for memristive devices to implement iSTDP and eSTDP learning functions \citep{talanov2017tr}. We have also presented the physical implementations and experimental results for dopamine modulation implemented as an amplification of learning impulses. Based on works of   \cite{hennequin_inhibitory_2017,vogels_inhibitory_2013,gurney_new_2015} we have extended our previous model with electronic schematic for bio-plausible neuromodulation for D1 and D2 dopamine receptors presented in the Fig. \ref{fig:block_diagram} (left graphs).

\subsection{Block diagram}

\begin{figure}[ht!]
  \centering
\includegraphics[width=1.0\textwidth]{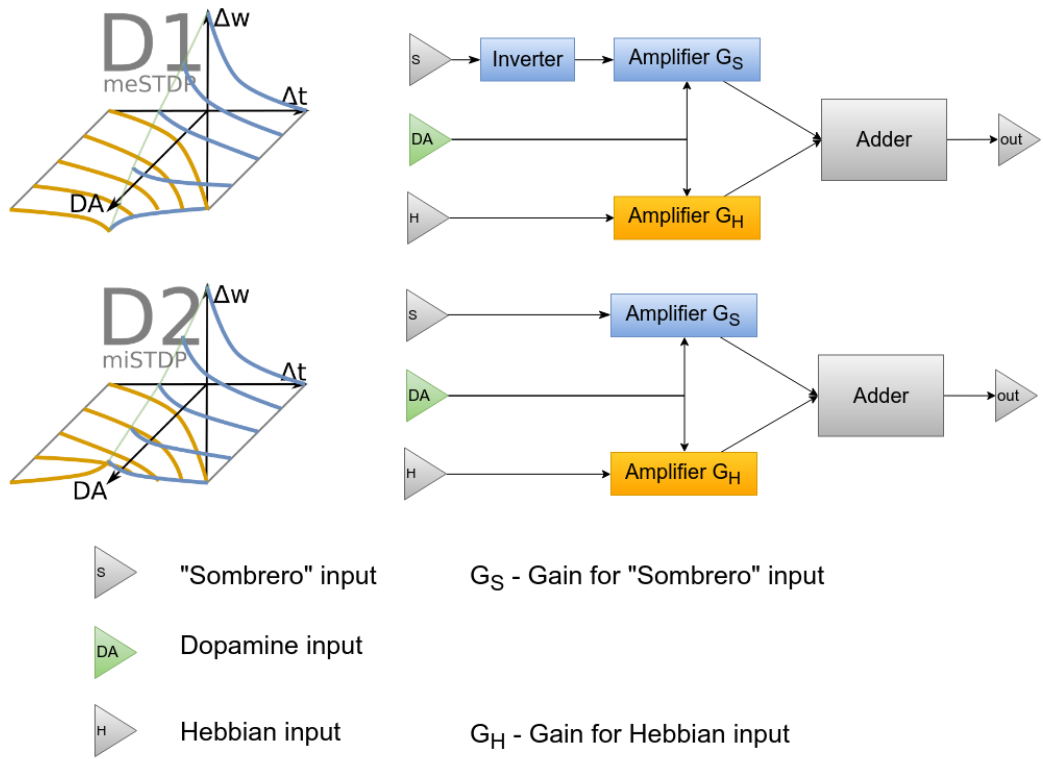}
\caption{The block diagram of the neuromodulatory feedback with inputs for: ``Sombrero'' impulses, Hebbian impulses and DA input that identifies the level of the dopamine. For the complete description of input functions please see our previous technical report \citep{talanov2017tr} }
\label{fig:block_diagram}
\end{figure}

We have decided to use mixture of two functions that represents extreme positions in the dopamine modulation STDP ``Sombrero'' and Hebbian indicated as $S$ and $H$ triangles in Fig. \ref{fig:block_diagram}. Input signals of ``Sombrero'' and Hebbian learning functions are balanced via dopamine level represented as $DA$ green triangle in Fig. \ref{fig:block_diagram}. The dopamine level setts up gain for each input "Sombrero" - $G_S$ and Hebbian - $G_H$ respectively in corresponding amplifier indicated as $Amplifier~G_H$ and $Amplifier~G_S$. The output of amplifiers is processed by the $Adder$ that implements the mixture of two modulated via dopamine input learning functions. Two cases of the dopamine modulation is represented as two panels in the Fig. \ref{fig:block_diagram} with the main difference in initial learning function for the low level of the dopamine: for D1 inverted -- ``Sombrero'' and for D2 -- ``Sombrero''. Thus the $Inverter$ that implements the inverted ``Sombrero'' for the low level of the dopamine is included in the block diagram for the D1 schematic.

\subsection{Wiring schematic}

\begin{figure}[ht!]
  \centering
\includegraphics[width=1.0\textwidth]{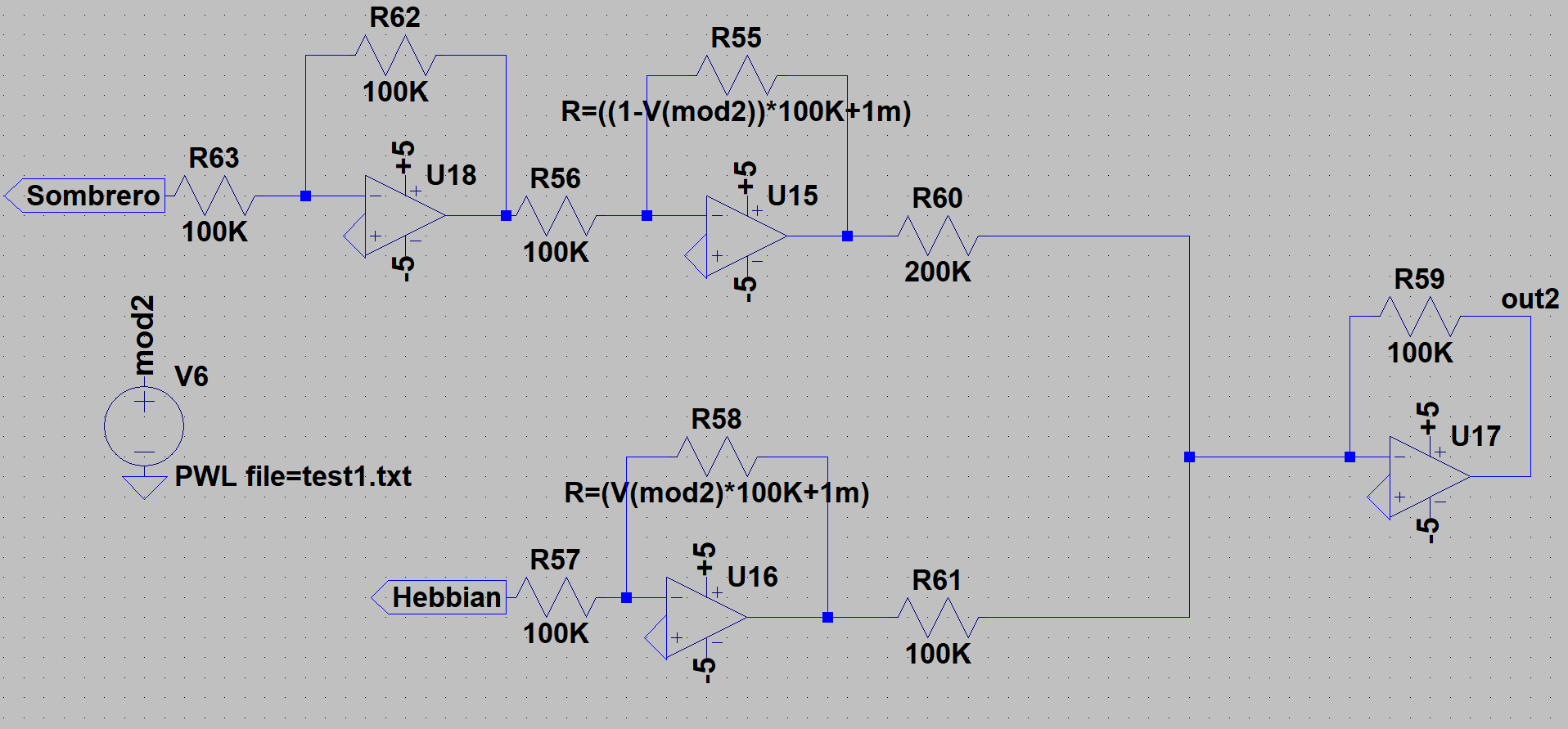}
\caption{The electronic circuit schematic of the D1 modulation. The $U18$ is inverter to implement the D1 extreme learning function the inverted ``Sombrero''. $U15$ and $U16$ are learning impulses inverting amplifiers with gain set by dopamine level indicated as $mod2$ that modulates the resistance of $R55$ and $R58$. The output weighted inverting adder is represented as $U17$, where ``Sombrero'' signal amplitude is twice lower than Hebbian.}
\label{fig:d1sch}
\end{figure}

\begin{figure}[ht!]
  \centering
\includegraphics[width=1.0\textwidth]{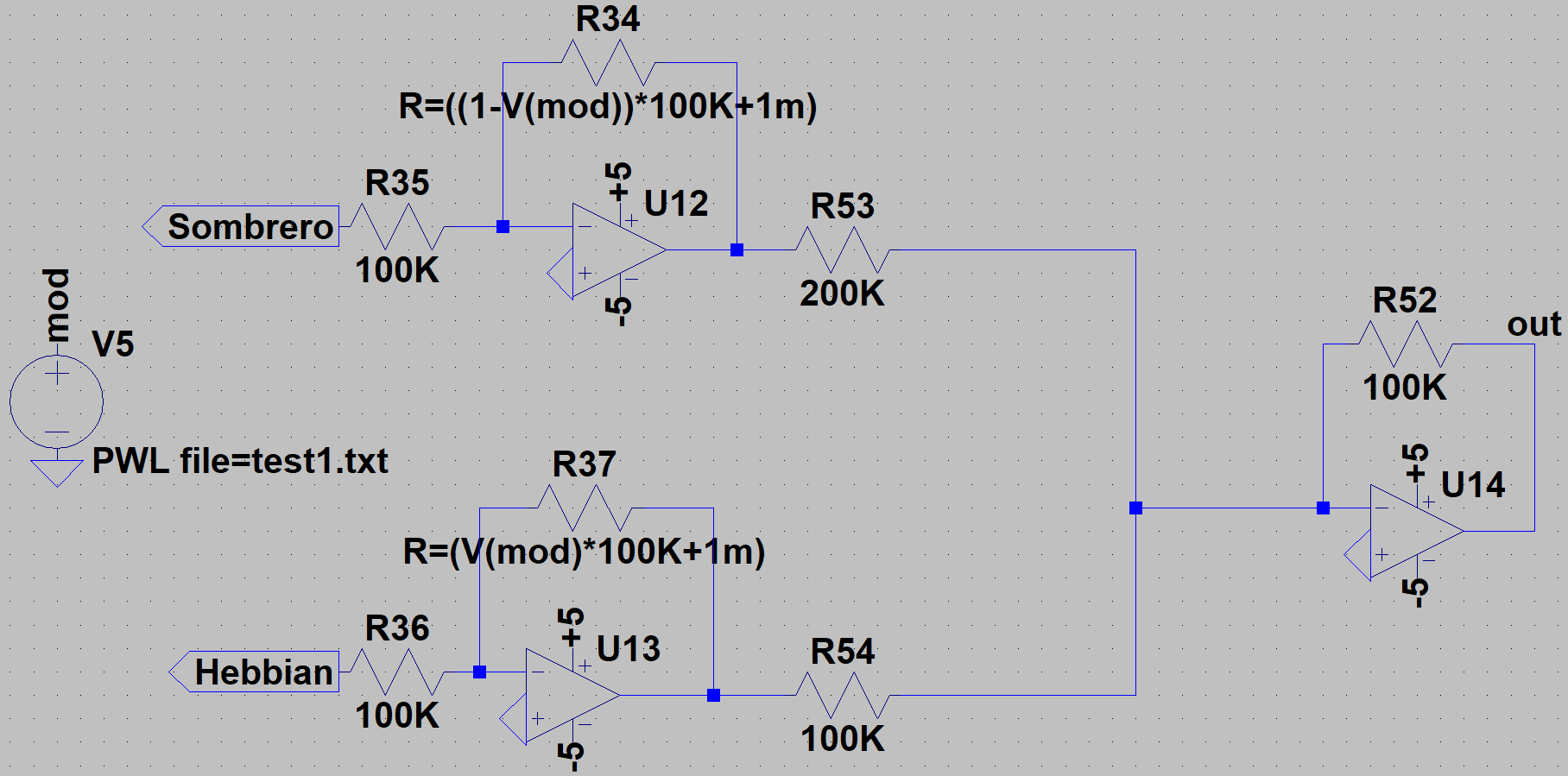}
\caption{The electronic circuit schematic of the D2 modulation. $U12$ and $U13$ are learning impulses inverting amplifiers with gain set by dopamine level indicated as $mod$ that modulates the resistance of $R34$ and $R37$. The output weighted inverting adder is represented as $U14$, where ``Sombrero'' signal amplitude is twice lower than Hebbian.}
\label{fig:d2sch}
\end{figure}

The wiring schematic of the dopamine modulation via receptor D1 is depicted in Fig. \ref{fig:d1sch}. The op-amp $U18$ plays the role of inverter for the ``Sombrero'' input signals to implement the extreme state of the learning function for D1 receptor the inverted ``Sombrero'' (see Fig.\ref{fig:block_diagram}). Two op-amps $U15$ and $U16$ are oppositely balanced amplifiers via resistances $R55$ and $R58$. The value of the $R55$ is set via digital potentiometer:
\begin{equation}
\label{eq:d1rs}
R=((1-V(mod2))*10^5+10^{-3})
\end{equation}

The value of the $R58$ is set via digital potentiometer opposed to $R55$:
\begin{equation}
\label{eq:d1rh}
R=(V(mod2)*10^5+10^{-3})
\end{equation}

When $V(mod)=1V$, according to eq.\ref{eq:d1rs}, $R55=10^{-3} Ohm$, which makes gain almost equal zero. At the same time, from eq. \ref{eq:d1rh} $R58=100K$, combined with $R57$ makes gain equals to one. If $V(mod)=0$ gains are opposite.
The outbound signal of both amplifiers $U15$ and $U16$ is processed by the weighted adder implemented via $U17$. The ``Sombrero'' amplitude is twice lower than Hebbian, to match global amplitude change in process of dopamine modulation.

The modulation schematic for D2 receptor is done in similar way except for the inverter of the ``Sombrero'' input signals that in case of D2 receptor is not used. The neuromodulatory schematic is represented in Fig. \ref{fig:d2sch}.

\subsection{Results}

\begin{figure}[htp]
  \centering
\includegraphics[width=1.0\textwidth]{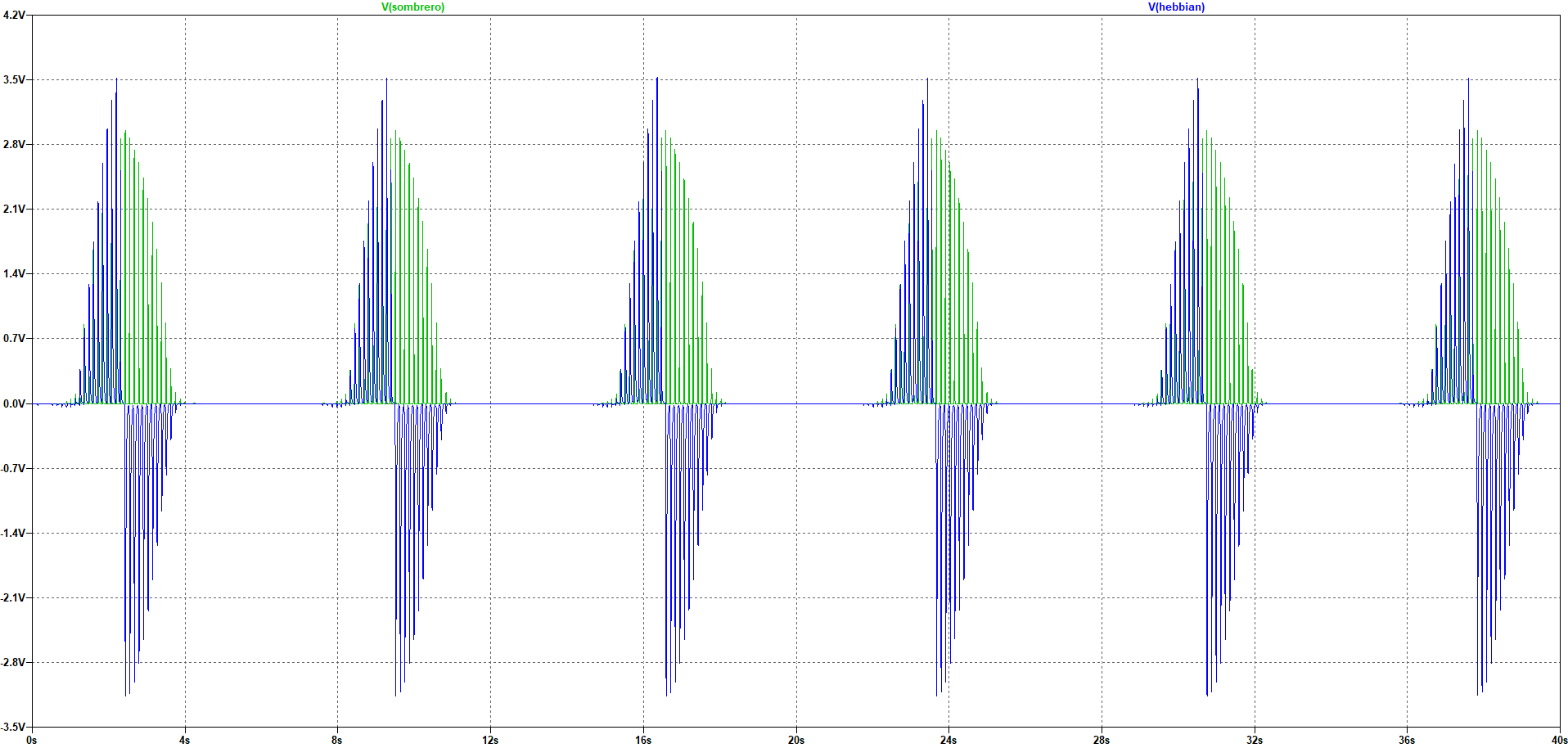}
\caption{``Sombrero'' (green) and Hebbian (blue) learning impulses.}
\label{fig:smb_hebb}
\end{figure}

\begin{figure}[htp]
  \centering
\includegraphics[width=1.0\textwidth]{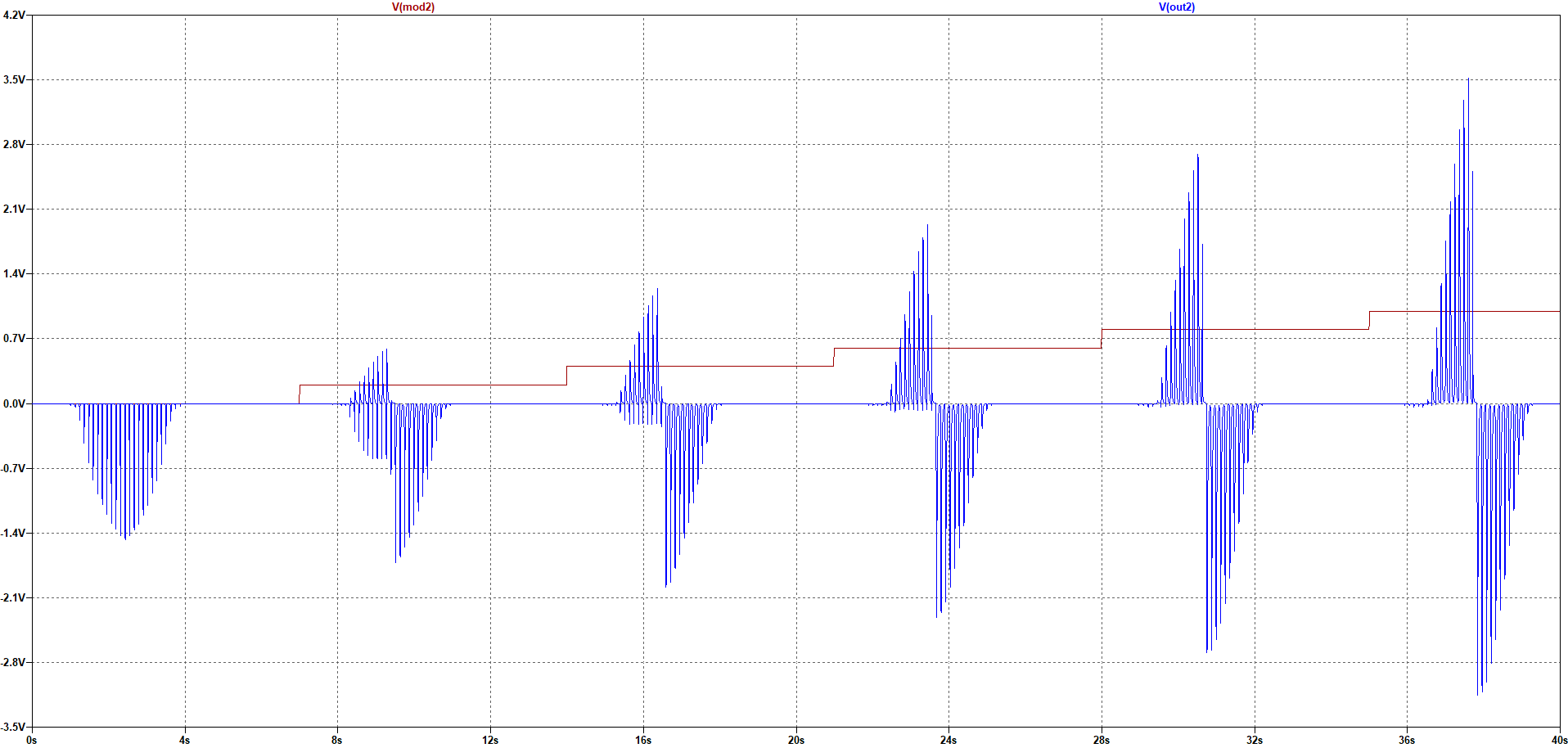}
\caption{Simulation of the D1 circuit in Fig.\ref{fig:d1sch}. ``Dopamine'' level (red) and D1 modulation result (blue).}
\label{fig:d1res}
\end{figure}

\begin{figure}[ht!]
  \centering
\includegraphics[width=1.0\textwidth]{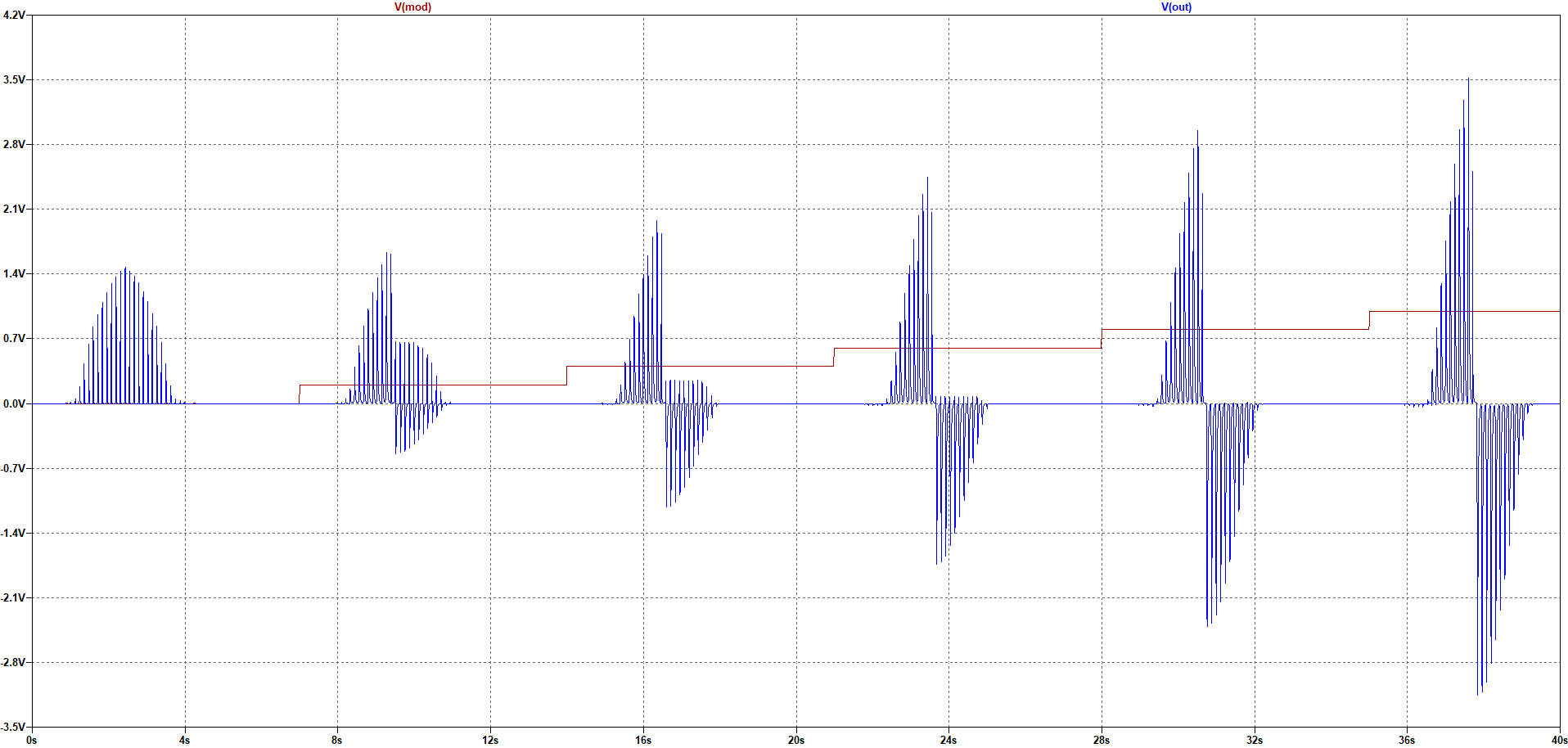}
\caption{Simulation of the D2 circuit in Fig.\ref{fig:d2sch}. ``Dopamine'' level (red) and D2 modulation result (blue).}
\label{fig:d2res}
\end{figure}

For the simulatory validation we have used LTSpice software framework. Inbound signals are presented in Fig. \ref{fig:smb_hebb} where green graph is the ``Sombrero'' signals and blue is the Hebbian signals for both the D1 and D2 dopamine receptors.

The results of D1 modulation are indicated in Fig. \ref{fig:d1res} where the proportion of the influence of the inbound learning function over outbound signal is defined by the level of the dopamine or set up of resistors $R55$ and $R58$. The level of the dopamine is represented as the red graph. Learning impulses that indicate the mixture of two learning functions inverted ``Sombrero'' and Hebbian are represented as blue graph.
The results of D2 modulation are presented in the Fig. \ref{fig:d2res} with the same color coding, where the level of the dopamine is identified via resistors $R34$ and $R37$.

The initial state of both graphs is low level of the dopamine and the learning function is inverted ``Sombrero'' for D1 and ``Sombrero'' for D2 respectively. The learning function for high level of the dopamine is Hebbian or $\frac{1}{x}$.
The gradual mixture of initial learning functions with final Hebbian is represented during simulatory experiments and is depicted on both figures starting from 4 seconds till 40 seconds of the simulation.

\bibliographystyle{apalike}
\bibliography{Electronic_schematic_for_bio-plausible_DA_modulation}

\begin{thebibliography}{}

\bibitem[Gurney et~al., 2015]{gurney_new_2015}
Gurney, K.~N., Humphries, M.~D., and Redgrave, P. (2015).
\newblock A {New} {Framework} for {Cortico}-{Striatal} {Plasticity}:
  {Behavioural} {Theory} {Meets} {In} {Vitro} {Data} at the
  {Reinforcement}-{Action} {Interface}.
\newblock {\em PLoS Biology}, 13(1):e1002034.

\bibitem[Hennequin et~al., 2017]{hennequin_inhibitory_2017}
Hennequin, G., Agnes, E.~J., and Vogels, T.~P. (2017).
\newblock Inhibitory {Plasticity}: {Balance}, {Control}, and {Codependence}.
\newblock {\em Annual Review of Neuroscience}, 40(1).

\bibitem[Talanov et~al., 2017]{talanov2017tr}
Talanov, M., Zykov, E., Gerasimov, Y., Toschev, A., and Erokhin, V. (2017).
\newblock Dopamine modulation via memristive schematic.
\newblock {\em CoRR}, abs/1709.06325.

\bibitem[Vogels et~al., 2013]{vogels_inhibitory_2013}
Vogels, T.~P., Froemke, R.~C., Doyon, N., Gilson, M., Haas, J.~S., Liu, R.,
  Maffei, A., Miller, P., Wierenga, C., Woodin, M.~A., Zenke, F., and
  Sprekeler, H. (2013).
\newblock Inhibitory synaptic plasticity: spike timing-dependence and putative
  network function.
\newblock {\em Frontiers in Neural Circuits}, 7:119.

\end{thebibliography}
\end{document}